\documentstyle[12pt]{article}

\advance\textheight by -5cm 
\advance\textheight by -2.5cm 
\begin{document}
\begin{titlepage}
\begin{center}
December 2, 1993     \hfill    LBL-34931\\
\vskip .4in
{\large \bf  Cluster Structure of Disoriented Chiral \\
Condensates in Rapidity Distribution}\footnote{This work was
supported
by the Director, Office of Energy
Research, Office of High Energy and Nuclear Physics, Divisions of High
Energy Physics and Nuclear Physics
 of the U.S. Department of Energy under Contract
DE-AC03-76SF00098, and by the Natural Sciences and
Engineering Research Council of Canada.}
\vskip .5in
Zheng Huang\footnote{Electronic Address: huang@theorm.lbl.gov}
and Xin-Nian Wang\footnote{Electronic Address: xnwang@nsdssd.lbl.gov}\\
 $^{\dagger}${\em Theoretical Physics Group, Mailstop 50A-3115}\\
$^{\ddagger}${\em Nuclear Science Division, Mailstop 70A-3307\\
    Lawrence Berkeley Laboratory\\
      University of California\\
    Berkeley, California 94720, USA}\\
\end{center}

\vskip .4in
\begin{abstract}
We study the creation of disoriented chiral condensates with some
initial boundary conditions that may be expected in the relativistic heavy
ion collisions. The equations of motion in the linear $\sigma$-model are solved
numerically with and without a Lorentz-boost invariance. We suggest that a
distinct cluster structure of coherent pion production in the rapidity
distribution may emerge due to a quench and may be observed in experiments.
\end{abstract}
\end{titlepage}

\newpage
\renewcommand{\thepage}{\arabic{page}}
\setcounter{page}{2}

Recently much attention has been paid to an interesting proposal to observe a
disoriented chiral condensate (DCC) in high energy collisions
\cite{anselm,blaizot,bj,other}. Such events
can be signaled by a coherent pion emission along some
particular isospin direction in the collision domain.
A probability distribution
of neutral pions is characterized by a non-binomial function if all isospin
directions are equally probable,
\begin{eqnarray}
P(r)=\frac{1}{2}\frac{1}{\sqrt{r}},
\end{eqnarray}
where $r=n_{\pi^0}/(n_{\pi^0}+n_{\pi^{\pm}})$. This behavior may have
already been observed in the so-called Centauro events in cosmic ray collisions
\cite{centauro}. The central theoretical work is to understand the
mechanism for
creating such a DCC domain in relativistic heavy ion collisions. Rajagopal and
Wilczek \cite{rw} have suggested that the expansion of highly relativistic
debris from collisions ``quenches'' the high temperature field
configurations such
that the long wavelength modes will grow with time, leading to a correlated
domain. It is recently reported, however, that the correlation size may be very
small \cite{ggp}.

In this paper, we shall study the creation of the disoriented chiral condensate
with a more realistic initial condition that may be expected in the high energy
collisions. We shall re-examine the idea of a quench based on a linear
$\sigma$-model and suggest that indeed, following a quench, correlated
pion fields will populate near the light cone where
the leading collision particles are
expanding at speed of light. The production of
the unusually rich (or poor) $\pi^0$'s will cluster
into groups in the rapidity distribution.
We are inspired by the consistency of our
results with some recent events of JACEE experiments \cite{jacee}
and confirm the general picture of a ``Baked Alaska'' sketched by
Bjorken, Kowalski and Taylor \cite{alaska}.

We assume that following a quench the possible DCC is to be described by
classical low energy effective interactions of pions at zero temperature
\begin{eqnarray}
{\cal L}=\int d^4x \left \{ \frac{1}{2}\partial_\mu\phi_i\partial^\mu \phi_i -
\frac{\lambda}{4}\left ( \phi_i\phi_i - v^2\right )^2 + H\sigma\right \}
\end{eqnarray}
where $\phi_i\equiv (\sigma ,\mbox{\boldmath $\pi$})$
stands for a vector in internal space.
$H\sigma$ is an explicit chiral symmetry-breaking term which is responsible for
the mass of the pseudo-goldstone bosons, the pions.
We look for a solution to the
classical equations of motion of $\phi$ with a given initial condition. Clearly
the initial condition is essential to determine
whether or not there will be DCC
in the system. For example, if the $\phi$ and $\dot{\phi}$ initially align in
the $\sigma$ direction throughout the space,
then the system does not have any pion
field at any time. What might be the typical initial configurations? In a
quench, the system begins at a temperature well above the chiral phase
transition point $T_c$. There are thermal fluctuations in all internal
directions. Therefore, it is appropriate to assume that initially the space
averages $\langle \phi\rangle\sim 0$ and
$\langle \dot{\phi}\rangle\sim 0$ but
$\langle \phi^2\rangle\neq 0$ and
$\langle \dot{\phi}^2\rangle\neq 0$ both for
$\sigma$ and $\pi$ fields. In addition,
we are modelling a situation applied to
highly relativistic collisions. The system should  satisfy the space-time
geometry of such a collision where the incident nuclei are, shortly after the
collision, highly Lorentz-contracted ``pancakes''
receding in opposite longitudinal
direction from the collision point \cite{bj83}. An approximate $1+1$ Lorentz
invariance of the system is indeed inspired by the existence of a
cental-plateau structure in the rapidity distribution of produced
particles in high energy cosmic-ray events\cite{jacee2} and
$pp$ or $p\bar{p}$ collisions \cite{cern}.
Of course,  the above picture should be modified
at large transverse distance, comparable to the nuclei radii.
However, in most of the transverse area,
the field changes rather slowly in the
transverse directions compared with the rapid
longitudinal expansion. Bearing
this approximation in mind, we shall assume a uniform field in
the transverse direction, i.e.\ $\phi$ is
only a function of $t$ and $z$. In terms
of more convenient coordinates, the proper time $\tau =\sqrt{t^2-z^2}$ and the
(spatial)rapidity variable
${\eta}=\frac{1}{2}\ln \frac{t+z}{t-z}$, the equations of motion read,
\begin{eqnarray}
\left [ \frac{1}{\tau}\frac{\partial}{\partial\tau}\left (\tau\frac{\partial}
{\partial\tau}\right )-
\frac{1}{\tau^2} \frac{\partial^2}{\partial {\eta}^2}\right ]
\sigma & = & -\lambda \sigma \left (\sigma^2+\mbox{\boldmath $\pi$}^2-
v^2\right )+ H\; ;
\label{a}\\
\left [ \frac{1}{\tau}\frac{\partial}{\partial\tau}\left (\tau\frac{\partial}
{\partial\tau}\right )-\frac{1}{\tau^2} \frac{\partial^2}{\partial
{\eta}^2}\right ]
\mbox{\boldmath $\pi$}& = & -\lambda \mbox{\boldmath $\pi$} \left
(\sigma^2+\mbox{\boldmath $\pi$}^2- v^2\right )\; .
\label{b}
\end{eqnarray}
Given an initial condition, Eqs. (\ref{a}) and (\ref{b})
completely determine the space-time evolution of the system.

The simplest case to solve Eqs. (\ref{a}) and (\ref{b}) is
when the system has an initial Lorentz-boost invariance, i.e.,
\begin{eqnarray}
\phi (\tau_0, {\eta})=\phi_0, \label{c}
\end{eqnarray}
where $\phi_0$ is independent of ${\eta}$. The lack of
$\frac{\partial}{\partial\tau}
\frac{\partial}{\partial {\eta}}$ in the Klein-Gordon
operator guarantees that Eq. (\ref{c}) is sufficient to
maintain the boost invariance at any time (there is no need to impose
$\frac{\partial}{\partial\tau}\phi (\tau_0, {\eta})={\rm constant}$).
To model a quench, we assume that the system is initially in a symmetric phase
and lies on the top of the ``Mexican Hat'' of the potential,
\begin{eqnarray}
\mbox{\boldmath $\pi$}(\tau_0)=0\quad ,
\quad \sigma (\tau _0)\simeq 0. \label{d}
\end{eqnarray}
To obtain a non-trivial solution for $\pi$ field, we allow a small initial
``kick'' of the system, i.e.\ a non-zero velocity $\frac{\partial
\phi}{\partial\tau}(\tau_0)\neq 0$. In this case, since the system starts from
one point $z\sim 0$ at $t\sim 0$ (the rapidity dependence of $\phi$ decouples
and the motion is zero-dimensional), the correlation in isospin direction is
automatically $100\%$. That is, if the initial kick is in $\pi^0$-direction,
then $\pi^{\pm}=0$ at all time such that there will be only $\pi^0$
production. A numerical solution is plotted in Fig.\ 1
 where we have used the following standard parameter input:
$v=87.4 \; {\rm MeV}$, $H=(119\;
{\rm MeV})^3$ and $\lambda =19.97$ so that
$f_\pi=92.5 \; {\rm MeV}$, $m_\pi=135 \; {\rm MeV}$ and
$m_\sigma= 600\; {\rm MeV}$. The initial condition we used in
Fig.~1 is $\phi(\tau_0)=0$,
$\partial\phi(\tau_0)/\partial\tau=(1,5,0,0)$ MeV/fm,
at $\tau_0=1$ fm/$c$.
The general feature of the solution is following.
The $\sigma$ field grows from zero and takes about
$1\; {\rm fm}/c$ proper time
to reach the true vacuum expectation value $\langle \sigma\rangle \sim f_\pi$,
while the $\pi$ field oscillates around zero rather slowly and eventually
tends to zero when the proper time gets large
$\tau =\tau_{\rm max}\sim 30\; {\rm fm}/c$.  In terms of
the real space-time variables $t$ and $z$, the disoriented chiral condensate
develops in the interior region of the receding pancakes, surrounded by the
normal vacuum in the exterior; as time ($t$) evolves, the DCC
will be squeezed toward the light cone, leaving behind a normal vacuum where
$\langle \mbox{\boldmath $\pi$}\rangle =0$ in the central
region inside the light cone.
For $t\gg \tau_{\rm max}$, the $\pi$ field is found only in a small region
$\Delta z\simeq \frac{\tau_{\rm max}^2}{2t}$ near the light front while the
$\sigma$ field occupies most of the space inside
the light cone, surrounded by a thin shell of DCC. This picture
supports a space-time geometry of DCC sketched in \cite{alaska}.

While the essential feature of the above solution may be generic, the real
situation is more complicated. One complication is
that the system may not start
from one point shortly after the collision and there may not be a Lorentz-boost
invariance in the system. In this case, the initial
condition like Eq.~(\ref{c}) is
not appropriate. If the system initially
has a distribution in rapidities and is
not correlated in isospin directions, the the crucial question would be whether
a quench can yield a coherent pion field at a later time.

We assume that shortly after the collision, the typical
configuration is that of a thermal random fluctuation  at high
temperature. The fluctuation is governed by the temperature.
Similarly to Ref.~\cite{rw}, we choose $\phi(\tau_0)$ and
$\partial\phi(\tau_0)/\partial\tau$ randomly according
to gaussian distributions so that
\begin{eqnarray}
\langle \phi (\tau_0, {\eta})\rangle _{\eta} =
\langle \frac{\partial}{\partial \tau}\phi (\tau_0, {\eta})\rangle _{\eta} =0
\label{f0}
\end{eqnarray}
but $\langle \phi^2 (\tau_0, {\eta})\rangle _{\eta}=v^2/4$ and
$\langle (\frac{\partial}{\partial \tau}
\phi(\tau_0, {\eta}))^2\rangle _{\eta}=v^2$.
By solving equations (\ref{a}) and (\ref{b}),
we find that as the proper time
evolves, the average value $\langle
\sigma \rangle _\eta$ grows from zero quickly and takes its
final value (unlike the case with boost-invariant initial condition,
the oscillation of $\langle\sigma \rangle _\eta$  around $f_\pi$
is quickly damped at large $\tau$);
 $\langle \mbox{\boldmath $\pi$}\rangle _{\eta}$
grows from zero and
oscillates around zero slowly in  the same way as in Fig.~1. Again,
$\langle \mbox{\boldmath $\pi$}\rangle _{\eta}$ decreases to zero
at about $\tau\simeq 20\sim 30\; {\rm fm}/c$. This
clearly indicates a strong correlation among the isospin directions in
rapidities. Indeed, we can define a correlation function at a given $\tau$,
\begin{eqnarray}
C(\tau, {\eta}-{\eta}' )=\frac{\mbox{\boldmath $\pi$}(\tau ,{\eta})\cdot
\mbox{\boldmath $\pi$}(\tau ,{\eta}' )}
{\mbox{\boldmath $\pi$}^2(\tau ,{\eta})+ \mbox{\boldmath $\pi$}^2(\tau ,{\eta}'
)}\; .
\end{eqnarray}
We find that after about $1\; {\rm fm}/c$
evolution in proper time (which is the time
scale that it takes for the average value of $\sigma$
field to reach its final value
$\langle \sigma\rangle _{\eta}\sim f_\pi$),
$C(\tau, {\eta}-{\eta}' )$ changes from a zero value
corresponding to a typical random distribution
to an exponential distribution with non-vanishing width as shown in Fig.~2.
For small $\Delta {\eta}={\eta}-{\eta}' $,
$C(\tau, \Delta {\eta})$ can be fit by an exponential function
 \begin{eqnarray}
C(\tau, \Delta {\eta})\propto \exp (-c_0m_\pi\tau |\Delta {\eta}|)
\quad\quad\quad
(\tau -\tau_0\geq 1\; {\rm fm}/c) \label{f}
\end{eqnarray}
where $c_0\sim 0.4$. At $\tau > 20\sim 30\; {\rm fm}/c$,
the correlation disappears as
indicated by the average value
$\langle \mbox{\boldmath $\pi$}\rangle _{\eta}$.
A similar result  is derived in \cite{1+1}
based on a non-linear $\sigma$-model
in $1+1$ dimensions where the pion fields are the phases of the order
parameter. However, in our model, we find that if initially
$\langle \phi\rangle^2_{\eta}\simeq f^2_\pi$, a correlated
distribution does not occur. This clearly
shows that the emergence of a correlation in
isospin orientation is mainly due
to the unstable long wavelength modes of the pion fields when $\langle
\phi\rangle ^2 < f_\pi ^2$.
Our result unambiguously points to the success of
the ``quench''  mechanism for creating DCC
(i.e.\ a high temperature initial condition
plus zero temperature equations of motion).

Eq.~(\ref{f}) indicates that a rapidity interval in which
the pion field is correlated in isospin direction could
be  as large as $2\sim 3$.  Our conclusion coincides
with the recent cosmic ray experiment~\cite{jacee}
where the dominant neutral pions are found  in an interval
$\Delta \eta \sim 3$.  Eq.~(\ref{f}) also suggests
that the correlation occurs mainly in
the region near the light cone where $\tau$ is small.
At a given large $t$, a region of ordinary vacuum
where $\langle \mbox{\boldmath $\pi$}\rangle _{\eta}$
vanishes effectively is found inside a shell of
coherent pion field. However, the cluster structure
of pions radiated from the coherent pion field
may occur anywhere in the whole rapidity region.
We should also emphasize that the width of the
cluster size in our study depends on the initial
time $\tau_0$ when the quenching happens. For sufficiently
large initial time, the width can become very small.

Although the size of correlated DCC domains may be small in space,
especially at a large $t$ because of the contracting effect due to
the relativistic expansion, we conclude that the DCC production can be
distinctly observed  as a cluster structure in rapidities. What is less
clear is whether there is indeed a quench after the collision.
Strictly speaking, the quench condition is an idealization of the more complex
situation where the temperature relaxes to zero only gradually.
For example, the hydrodynamics suggests a temperature drop according to
$T=T_0 (\frac{\tau_0}{\tau} )^{1/3}$  \cite{bj83}.
In this case, equations (\ref{a}) and (\ref{b}) may not be appropriate.
Clearly, more theoretical work is needed to examine if DCC can be created in
the process of cooling.\\
\vspace{12pt}
\begin{center}
{\bf Acknowledgements}
\end {center}

We wish to thank M.\ Asakawa, J.\ Bjorken, M.\ Suzuki and Dandi Wu
for very useful discussions.
This work was supported by the Director, Office of Energy
Research, Office of High Energy and Nuclear Physics, Divisions of High
Energy Physics and Nuclear Physics of the U.S. Department
of Energy under Contract No.\
DE-AC03-76SF00098, and by the
 Natural Sciences and Engineering
Research Council of Canada.

\pagebreak

\pagebreak
\begin{center}
{\large \bf Figure Captions}
\end {center}
\vspace{.5in}

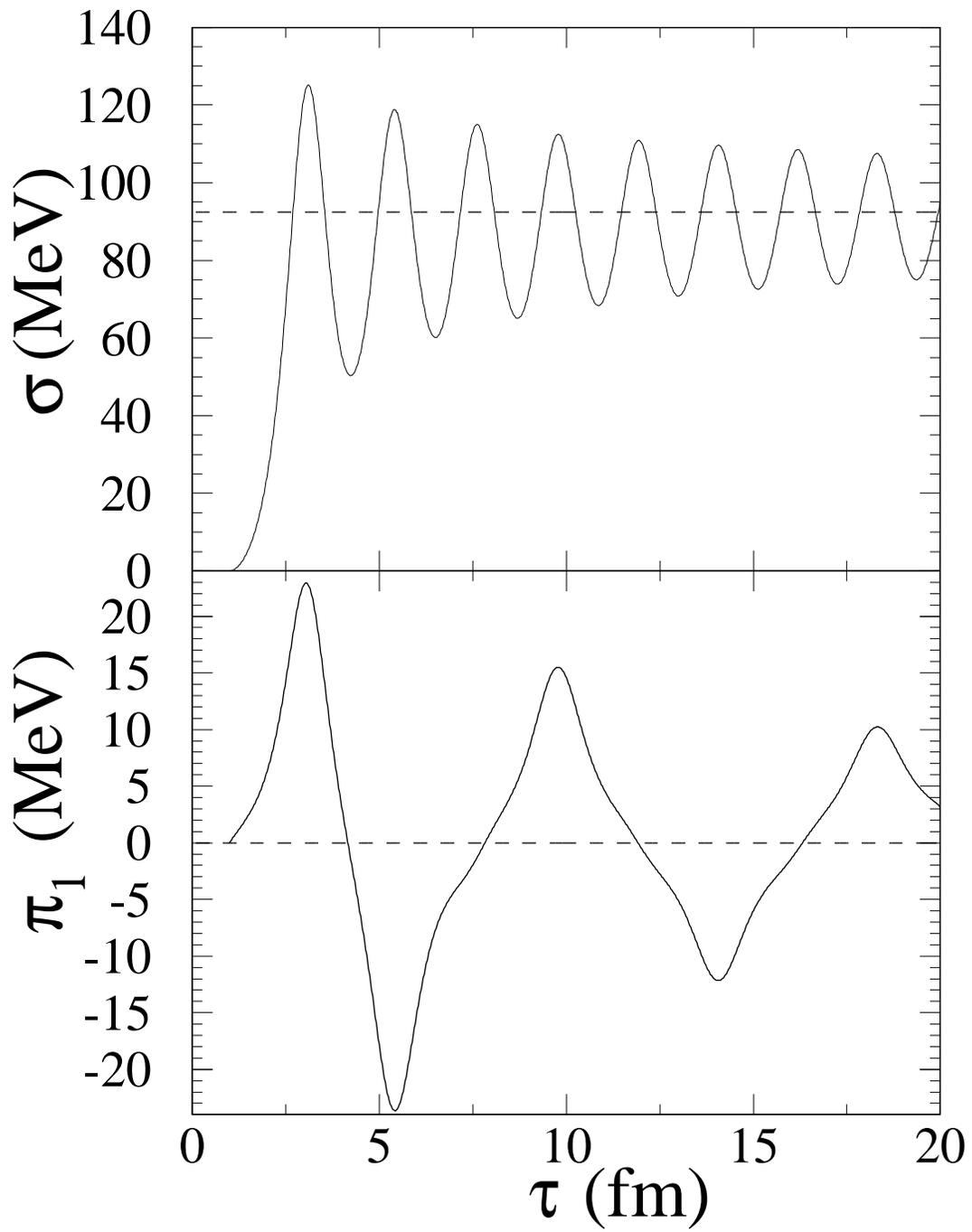
FIG.\ 1. Proper time evolution of $\sigma$ and $\pi_1$ fields following a
Lorentz-boost invariant  initial condition at $\tau_0=1$ fm/$c$.

\vspace{.6in}

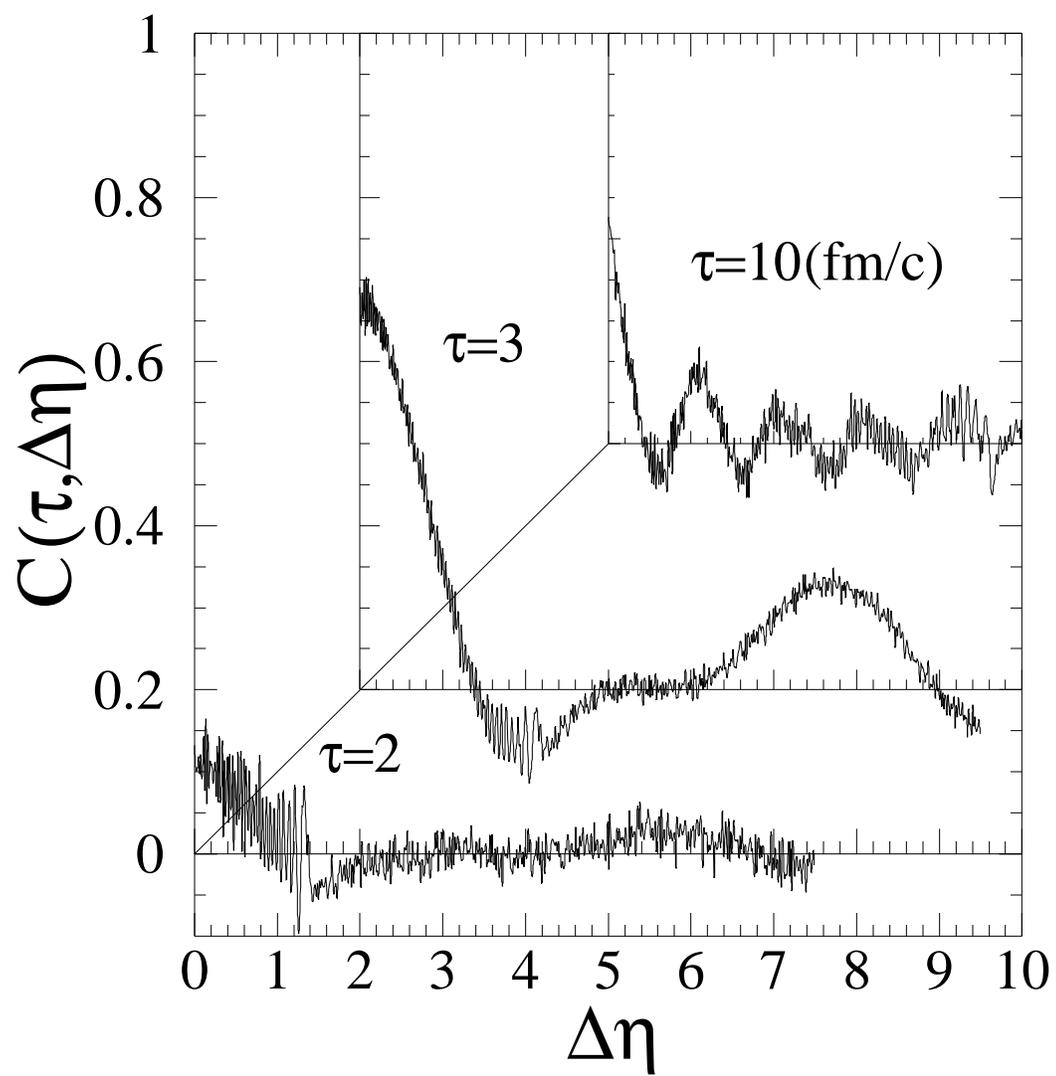
FIG.\ 2. The correlation function $C(\tau ,\Delta \eta)$ of DCC. The initial
configuration is that of a thermal fluctuation at $\tau_0=1$ fm/$c$.

\end{document}